\documentclass[aps,pra,showpacs,twocolumn,amsmath,amssymb,superscriptaddress,footinbib]{revtex4}

\usepackage[english]{babel}
\usepackage{times}
\usepackage{latexsym}

\usepackage{graphicx} 
\usepackage{epsf} 
\usepackage{subfigure} 
\usepackage{amsmath} 
\usepackage{amssymb} 
\usepackage{amsfonts}
\usepackage{bm}
\usepackage{natbib}
\usepackage{epstopdf}
\usepackage{appendix}

\def\be{\begin{equation}}
\def\ee{\end{equation}}
\def\bea{\begin{eqnarray}}          
\def\eea{\end{eqnarray}}
\def\bi{\begin{itemize}}
\def\ei{\end{itemize}}
\def\bin{\begin{enumerate}}
\def\ein{\end{enumerate}}

\def\bs#1{\boldsymbol{#1}}
\def\txt#1{\textrm{#1}}

\begin{document}

\title{Ultracold atomic gases in non-Abelian  gauge potentials: The case  of constant Wilson loop}


\author{
N. Goldman
}
\affiliation{Center for Nonlinear Phenomena and Complex Systems - Universit$\acute{e}$ Libre de Bruxelles (U.L.B.), Code Postal 231, Campus Plaine, B-1050 Brussels, Belgium}

\author{A. Kubasiak 
}
\affiliation{
ICFO-Institut de Ci\`encies Fot\`oniques,
Parc Mediterrani de la Tecnologia,
E-08860 Castelldefels (Barcelona), Spain}

\affiliation{Marian Smoluchowski Institute of Physics Jagiellonian University, Reymonta 4, 30059 Krak\'ow, Polska
}

\author{
P. Gaspard
}
\affiliation{Center for Nonlinear Phenomena and Complex Systems - Universit$\acute{e}$ Libre de Bruxelles (U.L.B.), Code Postal 231, Campus Plaine, B-1050 Brussels, Belgium}

\author{ 
M. Lewenstein}

\affiliation{
ICFO-Institut de Ci\`encies Fot\`oniques,
Parc Mediterrani de la Tecnologia,
E-08860 Castelldefels (Barcelona), Spain}
\affiliation{
ICREA - Instituci\`o Catalana de Ricerca i Estudis Avan{\c c}ats, 08010 
Barcelona, Spain}

\date{\today}

\begin{abstract}
Nowadays it is experimentally feasible to create artificial, and in particular, non-Abelian
gauge potentials for ultracold atoms trapped in optical lattices. 
Motivated by this fact, we investigate the fundamental properties of an ultracold Fermi gas
in a  non-Abelian $U(2)$ gauge potential characterized by a \emph{constant} Wilson loop. Under this specific condition, the energy spectrum  exhibits a robust band structure with large gaps and reveals a new fractal figure. The transverse conductivity is related to topological invariants and is shown to be quantized when the Fermi energy lies inside a gap of the spectrum. We demonstrate that the analogue of the integer quantum Hall effect for neutral atoms survives the non-Abelian coupling and leads to a striking fractal phase diagram. Moreover, this coupling induces an anomalous Hall effect as observed in graphene.
\end{abstract}

\maketitle
\section{Introduction}
Ultracold atoms in optical lattices offer unprecedented possibilities of controlling quantum matter and mimicking the systems of condensed-matter and high-energy physics \cite{Bloch, ouradv}.  Particularly fascinating is the possibility to study ultracold atoms under the influence of strong artificial Abelian and non-Abelian  ``magnetic" fields. The experimental realization of artificial Abelian ``magnetic" fields, which reproduce the physics of electrons in strong magnetic fields, is currently achieved through diverse schemes: for atoms in a trap the simplest way is to rotate the trap \cite{ouradv,Ho2001}, while for atoms in optical lattices this can be accomplished by combining laser-assisted tunneling and lattice acceleration methods \cite{Jaksch,Mueller,Demler}, using laser methods employing dark states \cite{Ohberg,Ohberg2}, using two-photon dressing by laser fields \cite{spielman,spielman2},  by the means of lattice rotations \cite{Holland1,Polini2005,Holland2,Tung}, or, last but not least, by the immersion  of atoms in a 
lattice within a rotating Bose-Einstein condensate (BEC) \cite{imme}.            
Several phenomena were predicted to occur in these arrangements such as the Hofstadter ``butterfly" \cite{Hofstadter} and the ``Escher staircase" \cite{Mueller}
in single-particle spectra, vortex formation \cite{ouradv, Holland1, Goldman2}, quantum  Hall effects \cite{Demler,Palmer,Goldman1, Holland2}, as well as other 
quantum correlated liquids \cite{Hafezi}.
  
As shown by one of us in Ref. \cite{Osterloh} (for an alternative method employing dark states see Ref. \cite{juze}), it is simple to generalize the scheme of Jaksch and Zoller 
 for generating artificial Abelian ``magnetic" fields \cite{Jaksch} in order to mimic artificial non-Abelian ``magnetic" fields. 
To this aim we have to consider atoms with more internal states (``flavors"). The gauge potentials that can 
be realized using standard manipulations, such as laser-assisted tunneling and lattice acceleration, can have practically 
arbitrary  matrix form in the space of ``flavors". In such non-Abelian potentials, the single-particle spectrum generally  
depicts a complex structure termed by one of us Hofstadter ``moth" \cite{Osterloh}, 
which is characterized by numerous extremely small gaps. The  model of Ref. \cite{Osterloh} 
has stimulated further investigations, 
including  studies of  nontrivial quantum transport properties \cite{Clark}, as well as studies of 
the integer quantum Hall effect (IQHE) for cold atoms \cite{Goldman1}, spatial patterns in optical lattices 
\cite{Goldman2}, modifications of the Landau levels \cite{santos}, and quantum atom optics \cite{santos2,santos3}.

One should note,
however, that  the $U(2)$ gauge potentials proposed in Ref. \cite{Osterloh} and used in most of the following works are 
 characterized by \emph{non-constant} Wilson loops: atoms performing a loop around a plaquette undergo a unitary transformation which depends on one of the spatial coordinates. Although such gauge potentials are interesting {\it per se}, the features characterizing the 
Hofstadter ``moth"  result from this {\it spatial
 dependence} of the Wilson loop, rather than from their non-Abelian nature.
 Indeed, the Hofstadter ``moth"-like spectrum  may actually be found in the standard Abelian case 
with a Wilson loop proportional to  $x$ (see Fig. 1). 

Two of us have shown that cold fermionic atoms trapped in optical lattices and 
subjected to artificial ``magnetic" fields should exhibit an IQHE \cite{Goldman1}. If a static force 
is applied to atoms, for instance  by accelerating the lattice, the
transverse Hall conductivity gives the
relation between this external forcing and the transverse atomic
current through the lattice. It has been shown that this transverse conductivity is quantized, 
$\sigma _{xy}=-\frac{C}{h}$, where $C$ is an integer 
and $h$ is Planck's constant.  Note that this quantity can be easily measured
from density profiles, as shown recently by Umucalilar {\it et al.} \cite{Umu}.
 The quantization
of $\sigma _{xy}$ occurs, however, only if  the Fermi energy of the system is located  inside a gap of the single-particle spectrum. While the observation of the IQHE  
seems to be  experimentally feasible in Abelian ``magnetic" fields, it is hardly so in the deeply non-Abelian regime in which 
the gaps of the ``moth" become very small \cite{rem}. 

The question therefore arises whether the consideration of non-Abelian gauge potentials characterized by a \emph{constant} Wilson loop could stabilize the spectral gaps and guarantee the robustness of the IQHE in ultracold fermionic gases
and whether an anomalous IQHE, as observed in graphene, can exist in such systems.

\begin{center} 
\begin{figure}
{\scalebox{0.23}{\includegraphics{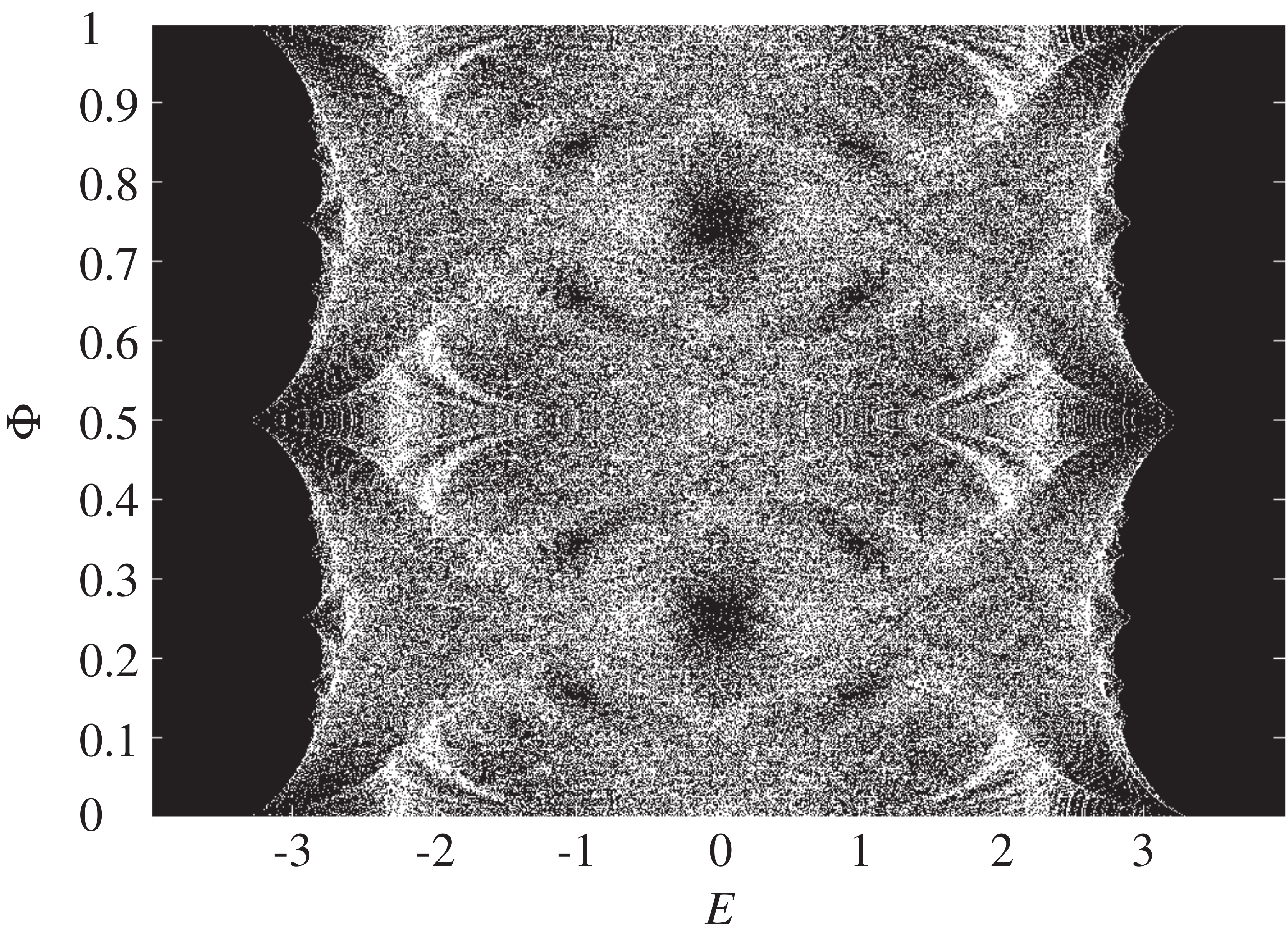}}} 
\caption{\label{abel} Energy spectrum $E=E(\Phi)$ in the case of the Abelian gauge potential $ \boldsymbol{A}=(0, 2 \pi \Phi m^2,0)$ with $x=ma$, corresponding to 
the \emph{non-constant} Wilson loop $W(m)=e^{i 2 \pi \Phi (1+2 m)}$.  Compare with the Hofstadter ``moth" depicted in Fig. 1 of Ref. \cite{Goldman1} or in Fig. 4 of Ref. \cite{Travel}. The energy $E$ is expressed in units of the tunneling amplitude $t$.} 
\end{figure} 
\end{center} 

In this work we provide affirmative answers to these questions by considering the IQHE in a system that features a  non-Abelian gauge potential characterized by  specific  non-commutating constant components and by a \emph{constant} Wilson loop. 
We calculate the energy spectrum and we obtain a robust
band structure with well developed gaps, which differs drastically from the case of the gauge potential of Ref. \cite{Osterloh}. In particular, we note the existence of van Hove singularities in the density of states and we obtain their analytical expression. We then evaluate the conductivity
$\sigma_{xy}$ for neutral currents using topological methods: we express $\sigma_{xy}$  in terms of the topologically invariant Chern numbers associated to each energy band \cite{remark}.  We eventually
present a salient fractal phase diagram which represents 
the integer values of the transverse conductivity inside the infinitely many
gaps of the spectrum. In this way, we show that the atomic transport is ruled by the \emph{quantum Hall effect supremacy}: the IQHE survives in  the non-Abelian regime, but undergoes strong modifications 
with striking similarity to the anomalous IQHE in graphene \cite{graphene}: the transverse conductivity suddenly changes sign due to the presence of van Hove singularities and is, under certain conditions, anomalous because of conical energy spectra.  

\section{Artificial gauge potentials in optical lattices}
We consider a system of non-interacting two-component fermionic atoms trapped in a 2D optical square lattice of unit 
length $a$, with sites at $(x=m a,y=n a)$, with $n,m$ integers. The non-interacting limit can be reached using Feshbach resonances, or simply at low densities. The optical potential is strong, so that  the tight-binding approximation holds. The Schr\"odinger equation for a single particle subjected to an artificial gauge potential then reads 
\begin{align}
&t_a (U_x \, \psi_{m+1,n}+  U_x^{\dagger} \, \psi_{m-1,n}) \notag \\
&+  t_b (U_y \, \psi_{m,n+1}+  U_y^{\dagger} \, \psi_{m,n-1})=E \, \psi_{m,n} ,
\label{ham}
\end{align}
where $U_x$ [resp. $U_y$] is the tunneling operator and $t_a$ [resp. $t_b$] is the tunneling amplitude in the $x$ [resp. $y$] direction. In the following, we use $a$ as the length, and $t_a=t_b=t$ as the energy units, and set $\hbar=c=e=1$, except otherwise stated. The tunneling operators are related to the gauge potential according to $U_x=e^{i A_x}$ [resp. $U_y=e^{i A_y}$].

Here we consider the gauge potential
\begin{equation}
\boldsymbol{A}= \bigl ( \alpha  \sigma_y , 2 \pi \Phi m +\beta \sigma_x , 0 \bigr ) ,
\label{gauge}
\end{equation}
where $\alpha$ and $\beta$ are parameters, $(\sigma_x$, $\sigma_y)$ are Pauli matrices
 and $\Phi$ is the number of Abelian magnetic flux quanta per unit cell. 
 
In order to realize such a potential one may consider the method of  Ref. \cite{Osterloh}. However, the specific form of this gauge potential allows us to consider an even more practical scheme based on a  generalization of the method  currently developed by Klein and Jaksch \cite{imme}, or the one developed by 
the NIST group \cite{spielman,spielman2}. A more detailed description of various proposals can be found in Appendix A.

The tunneling operators are $2 \times 2$ unitary matrices, 
\begin{align}
&U_x=\cos \alpha+i \sigma_y \sin \alpha , \notag \\
&U_y(m)=e^{i 2 \pi \Phi m} (\cos \beta +i \sigma_x \sin \beta ) ,
\label{eq3}
 \end{align}
 which act on the two-component wave function $\psi_{m,n}$. 

The single-particle Hamiltonian is invariant under translations defined by the 
operators $T^{q}_x \, \psi_{m,n}=\psi_{m+q,n}$ and $T_y \,  \psi_{m,n}=\psi_{m,n+1}$ under 
the condition that  $\Phi = \frac{p}{q}$, where $p$ and $q$ are integers.  Consequently, the system is 
restricted to a $q \times 1$ super-cell and one can express the wave function as 
$ \psi_{m,n}=e^{i k_x m} e^{i k_y n} u_m$ , with $u_m$ a $q$-periodic function. The wave 
vector $\boldsymbol{k}$ belongs to the first Brillouin zone, a $2$-torus defined as 
$k_x \in [0, \frac{2 \pi}{q}]$ and  $k_y \in [0,2 \pi]$. The Schr\"odinger equation \eqref{ham}  then reduces 
 to a generalized Harper equation
\begin{widetext}
\begin{align}
E \, u_m=& \begin{pmatrix}
\cos \alpha  & \sin \alpha \\ -\sin \alpha & \cos \alpha  \end{pmatrix} u_{m+1} e^{i k_x}   +  \begin{pmatrix}
\cos \alpha  & -\sin \alpha \\ \sin \alpha & \cos \alpha  \end{pmatrix} u_{m-1} e^{-i k_x}   \notag \\
&+ 2 \begin{pmatrix}
\cos (2 \pi \Phi m + k_y) \cos \beta   &  -\sin(2 \pi \Phi m + k_y) \sin \beta \\ - \sin (2 \pi \Phi m + k_y) \sin \beta &  \cos (2 \pi \Phi m + k_y) \cos \beta  \end{pmatrix} u_m .
\label{harper}
\end{align}
\end{widetext}

\section{Non-Abelian gauge potentials}

Artificial gauge potentials generally induce the following non-trivial unitary transformation for atoms hopping around a plaquette of the lattice:
\be
U=U_x U_y (m+1) U_x^{\dagger} U_y^{\dagger} (m).
\ee
 In the presence of the gauge potential  Eq. \eqref{gauge}, atoms performing a loop around a plaquette undergo the unitary transformation:
\begin{widetext}
\be
U= e^{i 2 \pi \Phi} \begin{pmatrix}
\cos ^2 \alpha + \cos 2 \beta \sin ^2 \alpha + \frac{i}{2} \sin 2 \alpha \sin 2 \beta 
& \sin 2\alpha \sin^2 \beta - i \sin^2 \alpha \sin 2\beta \\
-\sin 2\alpha \sin^2 \beta - i \sin^2 \alpha \sin 2\beta 
& \cos ^2 \alpha + \cos 2 \beta \sin ^2 \alpha - \frac{i}{2} \sin 2\alpha \sin 2\beta
\end{pmatrix}.
\label{Umatrix}
\ee
\end{widetext}
If one sets $\alpha=d \pi$ or  $\beta=d \pi$, where $d$ is an integer, the matrix  $U= \exp (i 2 \pi \Phi)$ is proportional to the identity and the system behaves similarly to the Hofstadter model \cite{Hofstadter}. When $\alpha=d \pi /2$ \emph{and} $\beta=d' \pi /2$, where $d$ and $d'$ are odd integers, one finds that $U= -\exp (i 2 \pi \Phi)$ and the system is equivalent to the $\pi$-flux model in which half a flux quanta is added in each plaquette \cite{graphene}. In these particular cases where $U= \pm \, e^{i 2 \pi \Phi}$, our problem decouples  into two independent Schr\"odinger equations describing particles  in an Abelian magnetic field: the system is in the Abelian regime. For any other values of the parameters $\alpha$ and $\beta$, the matrix $U$ is a non-trivial $U(2)$ matrix,  namely $U \not\propto \hat{1}$, and the system exhibits the non-Abelian Aharonov-Bohm effect: the system is in the non-Abelian regime. In Appendix B we present a more rigorous discussion of the gauge symmetry in our model, and explain in more details the meaning of the ``genuine Abelian" and ``genuine non-Abelian" cases. \\

\begin{center} 
\begin{figure}
{\scalebox{0.34}{\includegraphics{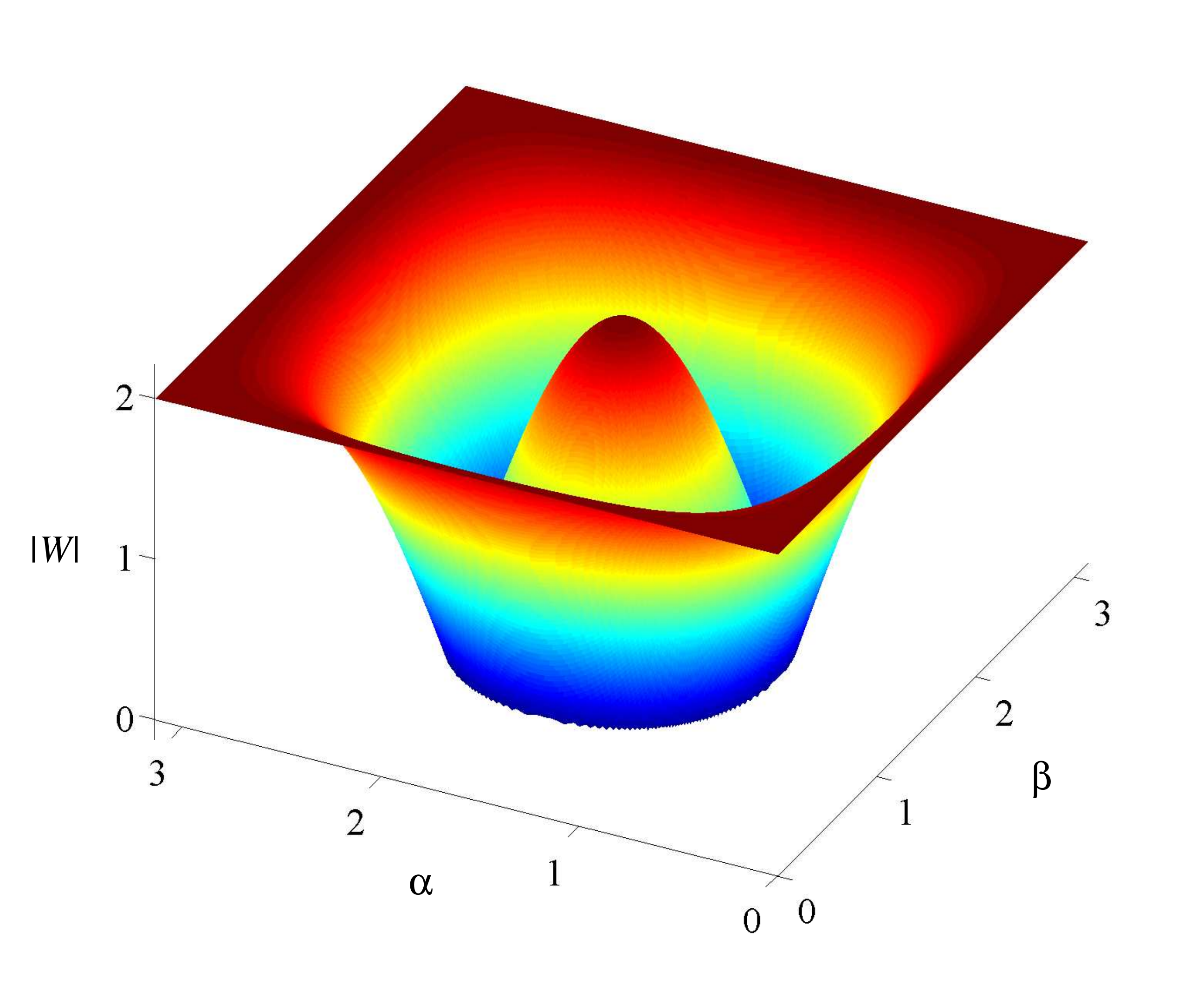}}} 
\caption{\label{wilson} (Color online) Wilson loop's magnitude as a function of the parameters $\vert W \vert= \vert W (\alpha , \beta) \vert$. The Abelian regime is determined by the criterion $\vert W \vert=2$: in the range $\alpha , \beta \in [0, \pi]$, the system is equivalent to the Abelian Hofstadter model along the lines $\alpha=0, \pi$ or  $\beta=0, \pi$, and is equivalent to the Abelian $\pi$-flux model  at the singular point  $\alpha=\beta=\pi /2$. For any other values of the parameters $\alpha , \beta \in [0, \pi]$, $\vert W \vert \ne 2$ and the system is non-Abelian in the sense that the unitary loop operator $U$ is not proportional to the identity matrix.} 
\end{figure} 
\end{center}

A criterion for distinguishing  between the Abelian and the non-Abelian cases is obtained by introducing the gauge invariant Wilson loop, 
\begin{align}
W&=\textrm{tr} \, U_x U_y (m+1) U_x^{\dagger} U_y^{\dagger} (m) \notag \\
&= 2 \, e^{i 2 \pi \Phi} \, (\cos ^2 \alpha + \cos 2 \beta \sin ^2 \alpha ) .
\end{align}
The unitary matrix $U$, Eq. \eqref{Umatrix}, is proportional to the $2 \times 2$ identity matrix if and only if $\vert W \vert =2$ (see Appendix B). Therefore the non-Abelian regime is reached under the condition that $\vert W \vert \ne 2$. Besides we note that the Wilson loop possesses the symmetry $W (\alpha , \beta)=W (\beta , \alpha )$. In Fig. \ref{wilson}, where we show the Wilson loop's magnitude as a function of the parameters, $\vert W \vert= \vert W (\alpha , \beta) \vert$,  we can easily identify the regions corresponding to the Abelian ($\vert W \vert = 2$) and to the non-Abelian regimes ($\vert W \vert \ne 2$). In this figure, the parameters evolve in the range $\alpha , \beta \in [0, \pi]$, and we note that the Abelian $\pi$-flux regime is reached at a singular point, $\alpha=\beta= \pi /2$.  

We also point out that the criterion
according to which the non-Abelian regime is reached when $[U_x,U_y] \ne 0$, and which can be found in previous works \cite{Osterloh,Goldman1}, doesn't apply in the present context: for the situation where $\alpha=d \pi /2$ \emph{and} $\beta=d' \pi /2$, where $d$ and $d'$ are odd integers, one finds that $[U_x,U_y] =2 i \, e^{2 i m \pi \Phi} \sigma_z$, while the system is Abelian because of its trivial Wilson loop, $\vert W \vert=\vert - 2 e^{2 i \pi \Phi} \vert =2$.

Contrary to the non-Abelian systems considered in previous works \cite{Osterloh,Clark,Goldman1}, we emphasize that the gauge potential Eq. \eqref{gauge} leads to a Wilson loop which does not depend on the spatial coordinates. In the following section, we show that this feature leads to energy spectra and fractal structures which significantly differ from the Hofstadter ``moth" \cite{Osterloh}.

\section{The energy spectrum}

The energy spectrum can be obtained through direct diagonalization of  Eq. \eqref{harper}. 

In the Abelian regime 
corresponding to $\alpha = d \pi$ or $\beta = d \pi$, where $d \in \mathbb{Z}$, one finds $q$ doubly-degenerated bands 
for $\Phi=\frac{p}{q}$. In this particular case, the representation of the spectrum  as a function of the flux $\Phi$ 
leads to the fractal Hofstadter ``butterfly" \cite{Hofstadter}. For the other Abelian case $\alpha=\beta=\frac{\pi}{2}$, the system behaves according to the $\pi$-flux lattice: the spectrum $E=E(\Phi)$ depicts a Hofstadter ``butterfly" which is contained between $\Phi=[0.5 ; 1.5]$, i. e. shifted by $\Phi=0.5$ with respect to the original ``butterfly", and the system remarkably describes zero-mass Dirac particles  \cite{graphene}.  

\begin{center} 
\begin{figure}
\begin{center} 
{\scalebox{0.08}{\includegraphics{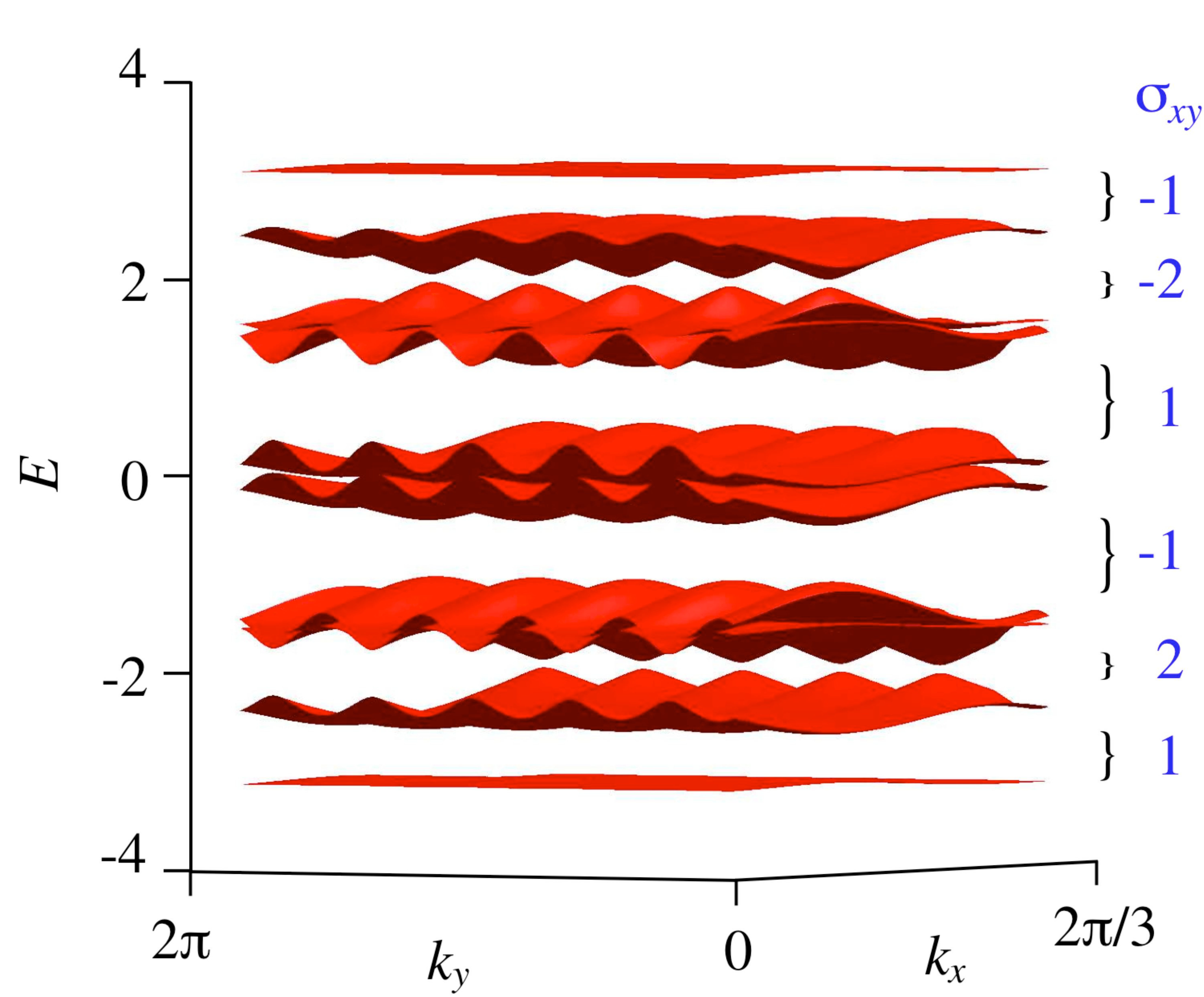}}} 
\caption{\label{bandfig} (Color online) Spectrum $E=E (k_x, k_y)$ for $\alpha=1$, $\beta=2$ and $\Phi=0.2$. While the degeneracy of some of the bands is lifted, the three central bands remain doubly-degenerate. Blue integers represent the quantized transverse conductivity $h \sigma_{xy}$ for Fermi energies situated inside the six gaps of the spectrum. The energy $E$ is expressed in units of the hopping parameter $t$. } 
\end{center}
\end{figure} 
\end{center}

\begin{center} 
\begin{figure}
\begin{center} 
{\scalebox{0.21}{\includegraphics{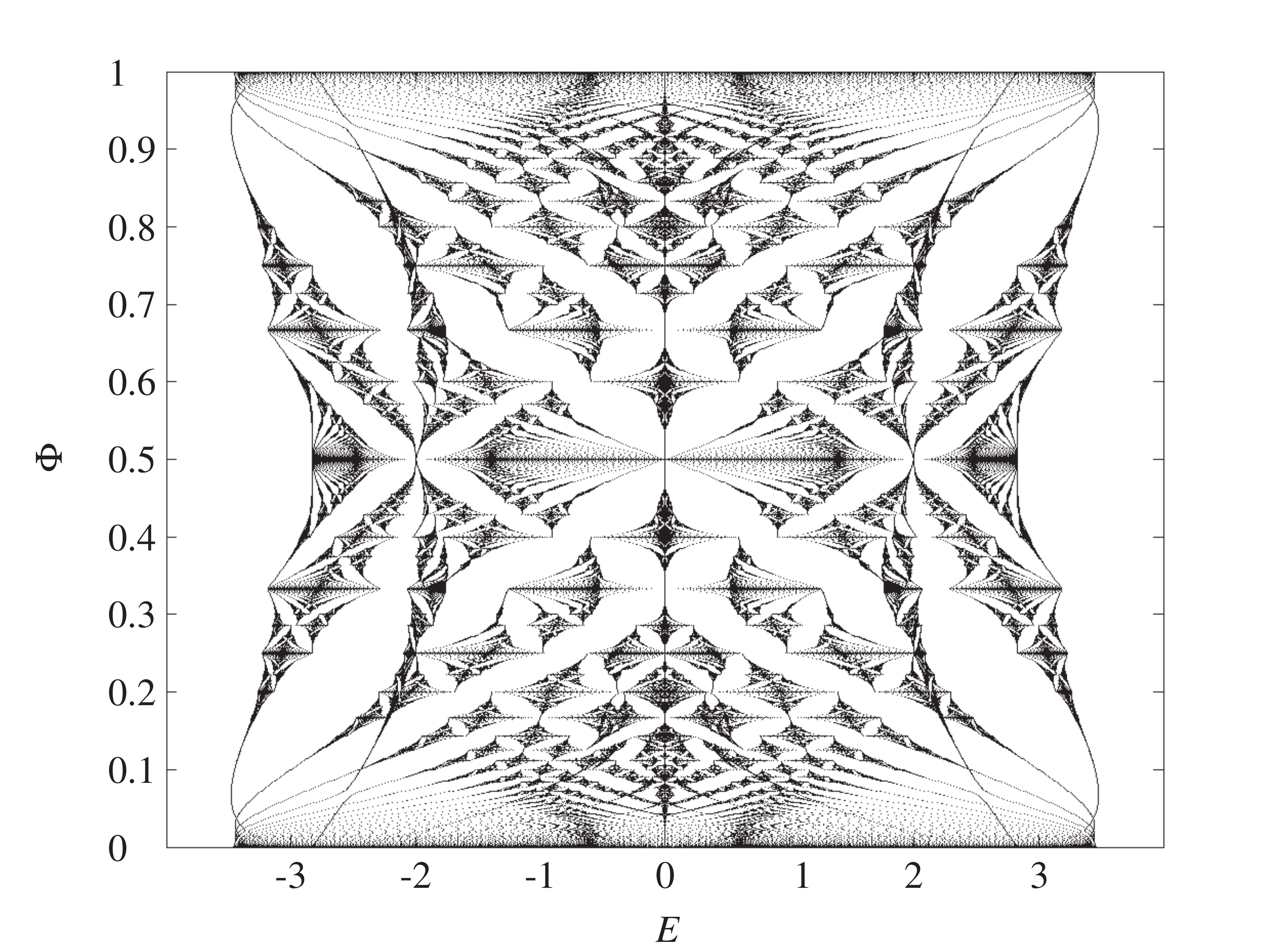}}} 
\caption{\label{fluxfig} Spectrum $E=E (\Phi)$ for $\alpha=\beta=\frac{\pi}{4}$ and $\Phi=\frac{p}{827}$, where $p$ is an integer. When $\Phi=0$, four van Hove singularities are located at $E= \pm (2 + \sqrt{2})$ and $E= \pm (2 - \sqrt{2})$. The energy is expressed in units of the hopping parameter $t$. } 
\end{center}
\end{figure} 
\end{center}

In the non-Abelian regime, which is 
reached for arbitrary values of the parameters $(\alpha, \beta)$, the degeneracy of certain bands is lifted and large gaps remain, as 
illustrated in Fig. \ref{bandfig}. For these general situations, the representation of the spectrum as a function of the flux $\Phi$  leads to new interesting features. As in the Abelian case, one observes repetitions of similar structures at various scales. However, new patterns arise in the non-Abelian case, as illustrated in Fig. \ref{fluxfig} for $\alpha=\beta=\frac{\pi}{4}$ and in Fig. \ref{one} for $\alpha=1$ and $\beta=2$. It is worth noticing that for arbitrary values of the 
 parameters $(\alpha, \beta)$, the spectra show well-developed gaps contrasting with the Hofstadter ``moth" which appears in the non-Abelian system proposed in Ref. \cite{Osterloh}.  We further notice that the spectrum is periodic with period $T_{\Phi}=1$ and is symmetric with respect to  $E=0$ and $\Phi=0.5$. 
 
 In the non-Abelian regime close to $\alpha, \beta =\pi /2$, one observes that  conical intersections are preserved in the energy spectrum. As shown in the next section, the particles behave similarly to Dirac particles in this non-Abelian region and the system exhibits an anomalous quantum Hall effect.
 
 We eventually note that when the flux $\Phi=0$, the density of states reveals several van Hove singularities at the energies $E= \pm 2 (1 + \cos \chi)$ and $E= \pm 2 (1 - \cos \chi)$, where $\chi=\alpha , \beta$. As the flux increases, these singularities evolve and generally merge. 
 
\section{Integer quantum Hall effect and the phase diagram}

We evaluate the linear response of the system described by Eq. \eqref{harper} to
an external force (lattice acceleration) applied along the $y$ direction and we evaluate the 
transverse conductivity $\sigma_{x y}$ using  Kubo's formula. Following the method of Ref. \cite{Goldman1}, one can generalize the well-known TKNN expression \cite{Thoules} to the present non-Abelian framework, yielding
\begin{align}
\sigma_{x y}&= \frac {1}{2 \pi  i h}  \sum_{E _{\lambda}
< E _{\rm F}}  \int_{\mathbb{T}^2} \sum_j \biggl ( \langle
\partial _{k_{x}} u_{\lambda j} \vert \partial _{k_{y}} u_{ \lambda
j} \rangle  \notag \\ 
&\qquad\qquad\qquad\qquad -\langle
\partial _{k_{y}} u_{\lambda j} \vert \partial _{k_{x}} u_{ \lambda
j} \rangle  \biggr ) d\boldsymbol{k} ,
\label{hall}
\end{align}  
where $u_{\lambda j}$ is the $j$th component of the wave function corresponding to the band $E_{\lambda}$ such that $H u_{\lambda}= E_{\lambda} u_{\lambda}$, and $\mathbb{T}^2$ refers to the first Brillouin zone of the system. The Fermi energy $E _{\rm F}$ is supposed to lie within a gap of the spectrum. The transverse conductivity  is then given by the contribution of all the states filling the bands $E _{\lambda}< E _{\rm F}$ situated below this gap. 

Eq. \eqref{hall} conceals a profound topological interpretation for the transverse conductivity  based on the fibre bundle theory \cite{Nakahara}. In the present framework, such bundles are conceived as the product of the parameter space $\mathbb{T}^2$ with the non-Abelian gauge group $U(2)$. This product space, which is supposed to be locally trivial but is generally expected to twist globally, is characterized by the non-Abelian Berry's curvature 
\be
\mathcal{F}=\bigl ( \partial_{k_x} \mathcal{A}^{y}-\partial_{k_y} \mathcal{A}^{x} + [\mathcal{A}^{x}, \mathcal{A}^{y}] \bigr ) dk_{x} dk_{y},
\ee
 where $(\mathcal{A}^{\mu})_{i j}=\langle u_{\lambda i} \vert  \partial _{k_{\mu}} u_{\lambda j} \rangle $ is the Berry's connection. The triviality of the fibre bundle is measured by the Chern number 
 \be
 C(E_{\lambda})= \frac{i}{2 \pi} \int_{\mathbb{T}^2} {\rm tr} \mathcal{F} ,
 \ee 
 which is a topological invariant and is necessarily an integer. Note that each band $E_{\lambda}$ is associated to a specific fibre bundle, on which a Chern number is defined. One eventually finds that the Hall-like conductivity Eq. \eqref{hall} is given by a sum of integer Chern numbers, 
\be
\sigma_{x y}= -\frac {1}{h}  \sum_{E _{\lambda}
< E _{\rm F}}  C(E_{\lambda}).
\label{final}
\ee
 As a consequence, the transverse Hall-like 
conductivity of the system evolves by steps corresponding to integer multiples of the 
inverse of Planck's constant and is robust against small perturbations.

The evaluation of these topological invariants leads to a complete understanding
of the IQHE which takes place in the present context. The aim is then to compute the 
Chern number associated to each band $E _{\lambda}$ of the spectrum. This computation can be achieved numerically thanks to an efficient method developed by  Fukui {\it et al.} \cite{Fukui} and which can be applied to our specific system. This method is summarized as follows: the Brillouin zone $\mathbb{T}^2$, defined by $k_x \in [0, \frac{2 \pi}{q}]$ and  $k_y \in [0, 2 \pi]$, is discretized into a lattice constituted by points denoted $\boldsymbol{k}_l=(k_{xl},k_{yl})$. On the lattice one defines a curvature $\mathcal{F}$ expressed as 
\be
\mathcal{F}_{12} (\boldsymbol{k}_l)= \textrm{ln} \, U_1 (\boldsymbol{k}_l) U_2 (\boldsymbol{k}_l +\hat{ \boldsymbol{1}}) U_1 (\boldsymbol{k}_l +\hat{ \boldsymbol{2}})^{-1} U_2 (\boldsymbol{k}_l)^{-1} , 
\ee
where the principal branch of the logarithm with $- \pi <  \mathcal{F}_{12}/i  \le \pi$ is taken,
$\hat{\boldsymbol{\mu}}$ is a unit vector in the direction $\mu$, and
\be
U_{\mu} (\boldsymbol{k}_l)= \sum_j \langle u_{\lambda j} (\boldsymbol{k}_l) \vert u_{\lambda j} (\boldsymbol{k}_l+\hat{ \boldsymbol{\mu}}) \rangle / \mathcal{N}_{\mu} (\boldsymbol{k}_l),
\ee
defines a link variable with a normalization factor $\mathcal{N}_{\mu} (\boldsymbol{k}_l)$ such that $\vert U_{\mu} (\boldsymbol{k}_l)\vert=1$. The Chern number associated to the band $E_{\lambda}$ is then defined by 
\be 
C=\frac{i}{2 \pi } \sum_l \mathcal{F}_{12} (\boldsymbol{k}_l).
\ee
 This method ensures the integral character of the Chern numbers and holds for non-overlapping bands. In the situations where the spectrum reveals band crossings, a more general definition of the link variables $U_{\mu} (\boldsymbol{k}_l)$ has been proposed in Ref. \cite{Fukui}. 
 
 We first compute the Chern numbers for the specific case illustrated in Fig. \ref{bandfig}. For $\alpha=1$, $\beta=2$ and $\Phi=0.2$, the Chern numbers associated to the seven bands are respectively $\{ -1 ; -1 ; 3 ; -2 ; 3 ; -1; -1 \}$. According to Eq. \eqref{final}, the transverse conductivity's values associated to the six gaps are $\{ 1 ; 2; -1 ; 1;  - 2 ; -1 \}$ as shown in Fig. \ref{bandfig} and in Fig. \ref{supp} a.

The phase diagram describing the IQHE for our model can eventually be drawn. 
 In this diagram we represent  the quantized transverse conductivity as a function of the Fermi 
energy $E _{\rm F}$ and flux $\Phi$. Here we illustrate a representative example of such a phase diagram 
which was obtained for $\alpha =1$, $\beta = 2$ (cf. Fig. \ref{one}). This striking figure differs radically 
from the phase diagrams obtained by Osadchy and Avron in the Abelian case \cite{Osadchy} since the Chern numbers associated to the gaps are no 
longer satisfying a simple Diophantine equation \cite{dioremark}. Consequently, the measurement of the transverse conductivity 
in this system should show a specific sequence of robust plateaus,  
heralding a new type of quantum Hall effect. \\
In order to give an alternative representation of this effect, we show two sections through this phase diagram, $\sigma_{xy}=\sigma_{xy} (E_F)$, at  $\Phi=0.2$ and $\Phi=0.02$ (see Fig. \ref{supp}).\\
This new effect is comparable to 
the IQHE observed in Si-MOSFET or  the anomalous IQHE observed in graphene
in the ``low flux" regime $\Phi \ll 1$ corresponding to experimentally available magnetic fields.
In this regime, the quantized conductivity evolves monotonically by steps of one between sudden changes of sign across the aforementioned van Hove singularities (see Fig. \ref{one} and Fig. \ref{supp} b). Moreover, in the vicinity of $\alpha , \beta = \pi/2$, the quantized conductivity increases by double integers because of Dirac points in the energy spectrum, in close similarity with the anomalous IQHE observed in graphene.

\begin{widetext}
\begin{center} 
\begin{figure}[h!]
{\scalebox{0.4}{\includegraphics{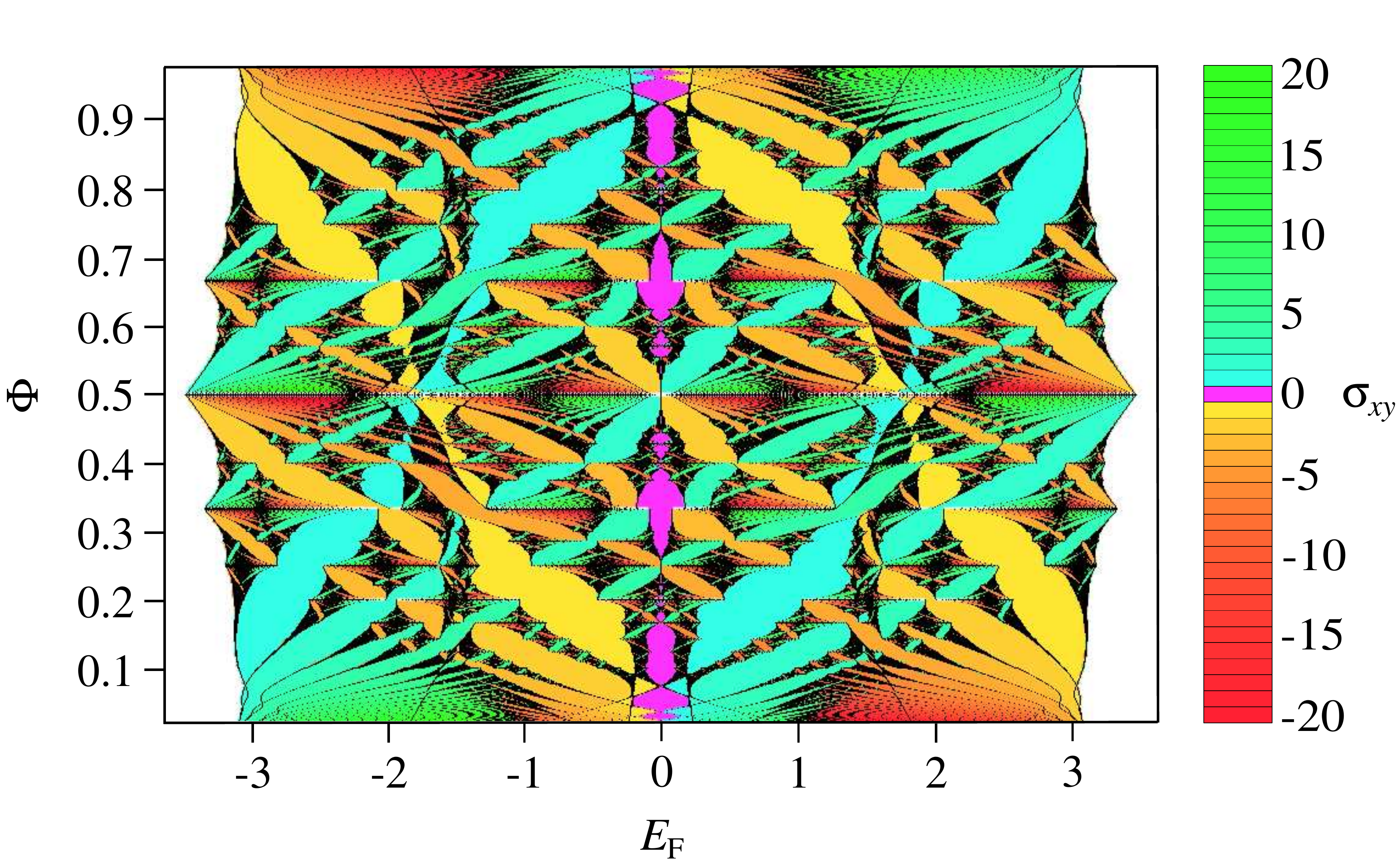}}} 
\caption{\label{one} (Color online) Spectrum $E=E(\Phi)$ and phase diagram for $\alpha =1, \beta = 2$ and $\Phi =\frac{p}{q}$ with $q<97$. 
Cold [resp. warm] colors correspond to positive [resp. negative] values of the quantized conductivity. 
Purple corresponds to a null transverse conductivity. For $\Phi \ll 1$, the quantized conductivity evolves monotonically but suddenly changes sign around the van Hove singularities located at $E \simeq \pm 1$ (see the alternation of cold and warm colors). The Fermi energy is expressed in units of the hopping parameter $t$ and the transverse conductivity is expressed in units of $1/h$. } 
\end{figure} 
\end{center} 
\end{widetext}

\begin{center} 
\begin{figure}
{\scalebox{0.29}{\includegraphics{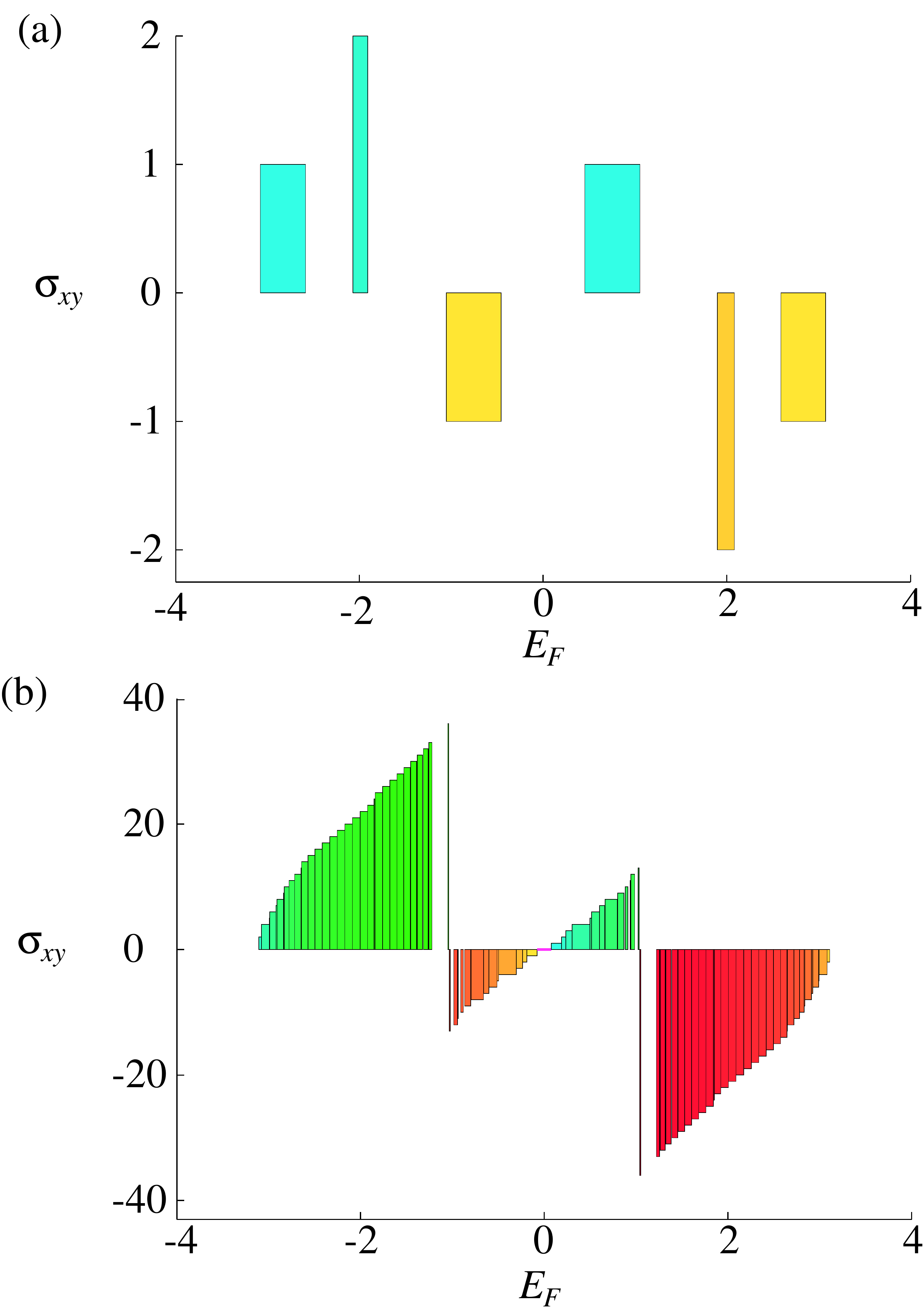}}} 
\caption{\label{supp} (Color online) Sections through the phase diagram illustrated in Fig. \ref{one}: $\sigma_{xy}=\sigma_{xy} (E_F)$, at $\Phi=0.2$ (a)  and  $\Phi=0.02$ (b). To each plateau $\sigma_{xy}=\textrm{constant}$, is associated a colored rectangle based on the line $\sigma_{xy}=0$. The color code and the parameters $(\alpha =1$, $\beta = 2)$ are the same as in Fig. \ref{one}. When $\Phi=0.2$ (a), one observes the sequence already represented in Fig. 3: the transverse conductivity associated to the six gaps is respectively $\{1;2;-1;1;-2;-1\}$. When $\Phi=0.02$ (b), the quantized conductivity evolves monotonically but suddenly changes sign around the van Hove singularities located at $E \simeq \pm 1$. The Fermi energy is expressed in units of the hopping parameter $t$ and the transverse conductivity is expressed in units of $1/h$.}
\end{figure} 
\end{center}

\section{Conclusions}

Summarizing, we have proposed in this paper how to realize in cold atomic systems a textbook example of 
 non-Abelian gauge potential characterized by a \emph{constant} Wilson loop. 
Our main result is that despite the coupling between the different ``flavor" components of the 
single-particle wave functions, the spectrum exhibits well-developed gaps of order of 0.1-0.2$t$, 
i.e. about 50-100 nK. 

The IQHE survives in the deeply non-Abelian regime and acquires a unique character 
specific to the non-Abelian nature of the gauge fields. It is characterized by a particular sequence of 
robust plateaus corresponding to the quantized values of the transverse conductivity. Moreover, the 
non-Abelian coupling induces controllable van Hove singularities as well as an anomalous Hall effect, similar to the effect induced by the hexagonal geometry in graphene. Experimental observation of this distinctive effect requires to achieve $T$ smaller than the gaps, 
i.e. of order of 10-50 nK, which is demanding but not impossible. 

The main experimental challenge 
consists here in combining several established methods into one experiment: 
laser assisted tunneling \cite{Jaksch}, BEC immersion \cite{imme}, and density profile measurements \cite{Umu}.

We acknowledge support of the EU IP Programme SCALA, ESF-MEC Euroquam Project FerMix, Spanish MEC grants (FIS 2005-04627, 
Conslider Ingenio 2010 ``QOIT), the Belgian Federal Government (IAP project ``NOSY"), and the "Communaut\'e fran\c caise de Belgique" (contract "ARC" No. 04/09-312) and the F.R.S.-FNRS Belgium. M.L acknowledges ERC AdG "QUAGATUA" and Alexander von Humboldt Stiftung. N. G. thanks S. Goldman for his comments, Pierre de Buyl for its support and ICFO for its hospitality. A.K. acknowledges support of the Polish Government Research Grant for 2006-2009 and thanks J. Korbicz and C. Menotti for discussions. The authors thank D. Jaksch, J. V. Porto, and C. Salomon for their valuable insights in various aspects of this work.

\appendices
\section{Experimental realizations}

The present work is devoted to the study of single-particle physics of cold atoms subjected to artificial gauge potentials. Recently several proposals have been made in order to realize such fields through diverse experimental techniques. For completeness we present in this appendix several feasible proposals for realizing Abelian and non-Abelian gauge potentials for cold atomic gases captured in a single trap or in an optical lattice.

\paragraph{\bf Abelian gauge fields} Here we list the most promising proposals for generating Abelian gauge potentials:

\begin{itemize}

\item{\it Rotating traps} Perhaps the simplest and most common way is to use rotating traps. Under rotation of the system, the centrifugal forces act as a Lorenz force for the atoms dynamics. 
In harmonic traps, when the rotation frequency approaches the trap frequency, the system expands and the atoms behave as an electronic system subjected to a constant magnetic field  directed along the rotation axis \cite{Bloch,ouradv,Ho2001}.
This method, which has been employed by many experimental groups, suffers from instabilities as the rotation frequency approaches the value of the trap frequency.

\item{\it Rotating lattices} In order to simulate an artificial magnetic field for particles evolving on a lattice, one has to superimpose a corotating lattice on a rotating Bose-Einstein condensate (BEC) \cite{Tung}. In such arragements, the quantum Hall effect and the physics of vortices can be subtlely explored \cite{Holland1, Polini2005, Holland2,Tung}.

\item{\it Laser methods employing dark states} Perhaps among the most promising, these methods can be used in single traps and in optical lattices. This method is based on the fact that the adiabatic motion of a $\lambda$-type 3-level atom creates a non-degenerate dark state in the presence of external laser fields. If such a dark state is space dependent, the atomic wave function aquires a topological (or Berry's) phase, which can be interpreted as the effect of an ``artificial" magnetic field \cite{Ohberg,Ohberg2}. 

\item{\it Laser assisted tunneling and lattice tilting} Several proposals have been made 
in order to realize ``artificial" magnetic fields in  optical lattices by using laser assisted tunneling \cite{Jaksch,Demler,Mueller}. The first method along thoses lines was proposed by Jaksch and Zoller \cite{Jaksch} and can be summarized as follows. The method uses two superimposed lattices: one lattice traps atoms in one internal state (red),  while the other traps atoms in another internal state (blue). Consequently, the entire lattice is realized in such a way that it consists of arrays of alternating columns (red-blue-red-blue-...) along the $y$ direction. One can use for this aim the states from different  hyperfine manifolds. Then by tilting the lattice in the $x$-direction, one creates a linear energy level shift. The tunneling along the $y$-direction is standard, while the tunneling along the $x$-direction in laser assisted - it occurs via a stimulated two-photon Raman transition tuned resonantly to the energy shift (constant for neighboring columns). The Raman laser has a global phase that depends on the $y$-coordinate, and which induces a phase to the tunneling matrix elements. Again, these phases modify the atoms dynamics and can  be interpreted as the effect of an ``artificial" magnetic field.

\item{\it BEC immersion} Very recently, a novel and elegant proposal has been made by Jaksch and Klein \cite{imme}. The method consists of immersing the optical lattice and the atoms of interest ($A$-atoms) into a BEC consisting of $B$-atoms. The $A$-atoms collide with $B$-atoms and create phonon excitations, which in turn will induce effective interactions between $A$-atoms \cite{imme}. If the immersion occurs into a \emph{rotating} BEC, the resulting effective interactions induce space-dependent phases to the tunneling matrix elements of the $A$-atoms. Therefore these phases modify the dynamics of $A$-atoms and simulate the presence of a magnetic field.

\item{\it Two-photon dressing by laser fields} Eventually one can use a two-photon dressing field in order to create an artificial vector potential, as demonstrated recently by Lin {\it et al.} \cite{spielman2} for Bose condensed $^{87}$Rb atoms in the $F=1$ manifold. The created dressed states are momentum and spin superpositions, and the effective Hamiltonian corresponds exactly to the one of charged particles in the presence of a ``magnetic" vector potential. The magnitude of this ``artifical" field is set by the strength and detuning of the dressing Raman fields. So far the method has allowed to generate a uniform vector potential (i.e. zero magnetic field), but the extension to a non-uniform potential with non-vanishing curl should be straightforward \cite{spielman2}. 

\end{itemize}

\paragraph{\bf Non-Abelian gauge fields} Some of the methods listed above and allowing for the realization of ``artificial" Abelian gauge potentials  can be extended in order to create non-Abelian fields. We list below the methods that can be considered in order to create a gauge potential of the form Eq.\eqref{gauge}, which is of interest in the present work. We focus on the possibilities to use alkali atoms with several internal states (which may be states from different and/or from the same hyperfine manifold).

\begin{itemize}

\item{\it Laser methods employing dark states} As pointed out in Ref. \cite{juze}, these methods can be generalized to the non-Abelian case, employing degenerate dark states. Such states arise in a four level system with three levels coupled to a common fourth  (the so-called ``tripod" configuration) under pairwise two-photon-resonance conditions. In particular, such dressing fields can be applied to atoms trapped in optical lattices.

\item{\it Laser assisted tunneling and lattice tilting} As mentioned above the method of Jaksch and Zoller can be directly generalized to the non-Abelian case \cite{Osterloh}. This can be acheived by considering atoms in two internal hyperfine manifolds (red and orange) and using Zeeman states as ``flavor" or ``color"  states. 
The case of $U(2)$ gauge fields is presented in Fig. \ref{schemat}.
The lattice is tilted in the $x$ and $y$ directions and a static magnetic field is used in order to split the Zeeman states energies. The tunneling in both horizontal and vertical directions is laser assisted, and the Raman lasers are tuned  in such a way that when an atom in a given Zeeman sublevel tunnels, it changes its state into a coherent superposition of all the Zeeman states. Controlling the Raman lasers strength and detuning  allows therefore to control the values of the parameters $\alpha$ and $\beta$ in the potential Eq.\eqref{gauge}. Additionally if the laser acting in the $y$ direction has a phase that depends linearly on
the $x$ coordinate, a non-vanishing 
Abelian gauge field, characterized by the flux $\Phi$, is created. 

\begin{center} 
\begin{figure}
{\scalebox{0.7}{\includegraphics{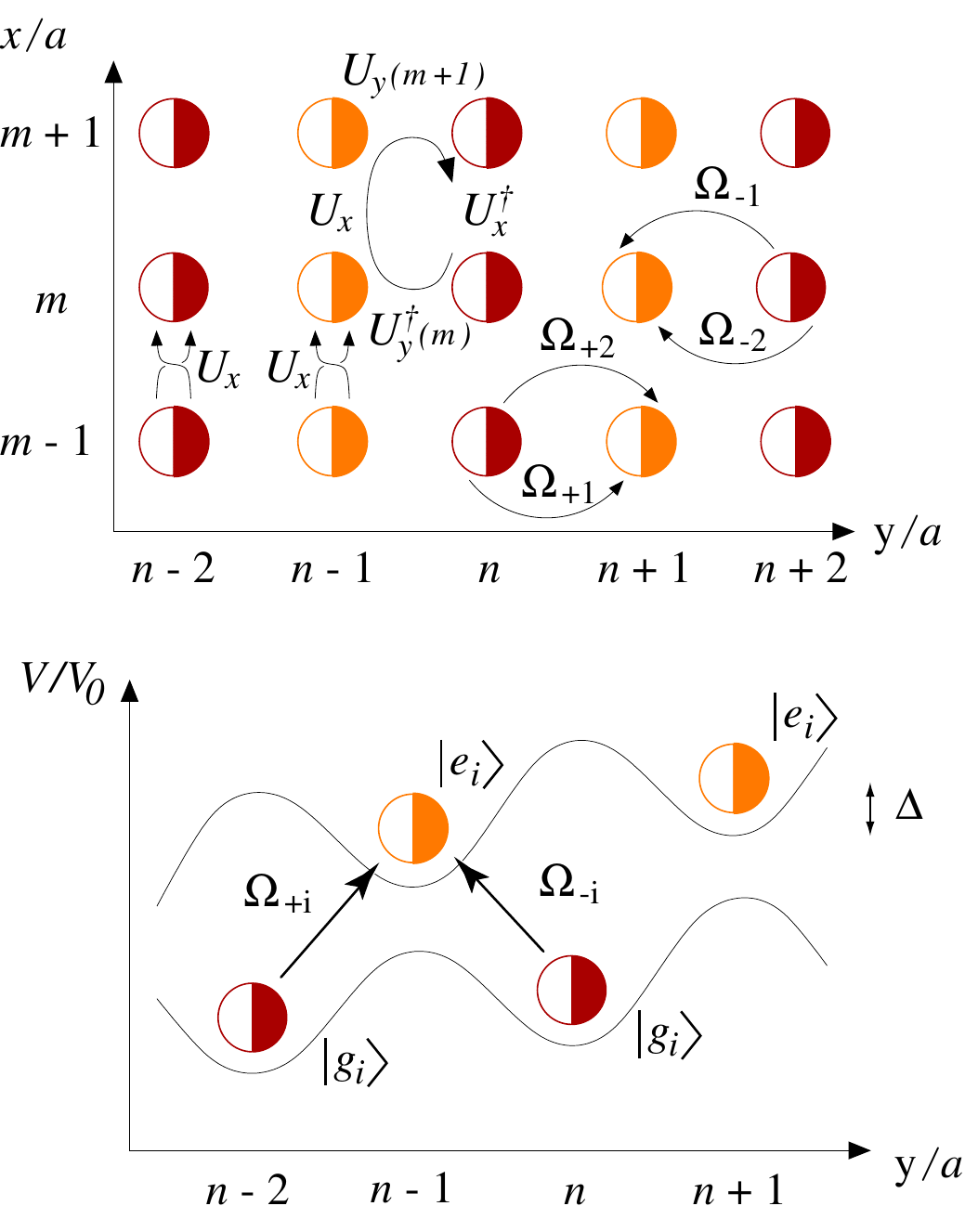}}} 
\caption{\label{schemat} (Color online)  \emph{Optical lattice setup for $U(2)$ gauge potentials}: Dark red empty semicircles [resp. dark red filled semicircles] denote atoms in states $\vert g_1 \rangle$ [ resp. $\vert g_2 \rangle$] and light orange empty semicircles [resp. light orange filled semicircles] denote atoms in states $\vert e_1 \rangle$ [ resp. $\vert e_2 \rangle$]. Top: Hopping in the $x$ direction is laser assisted and allows for unitary exchange of colors; it is described by the same unitary hopping matrix $U_x$ for both $\vert g_i \rangle$ and $\vert e_i \rangle$ states. Hopping along the $y$ direction, which is described by the hopping matrix $U_y(m)$, is also laser assisted and attains $m$-dependent phase factors. Bottom: Trapping potential in the $y$ direction. Adjacent sites are set off by an energy $\Delta$ due to the lattice acceleration, or a static inhomogeneous electric field. The lasers $\Omega_{\pm i}$ are resonant for transitions $\vert g_i \rangle \rightarrow \vert e_i \rangle$ for $n \rightarrow n \pm 1$. Because of the spatial dependence of $\Omega_{\pm i}$ (running waves in the $\pm x$ direction), the atoms hopping around the plaquette get the unitary transformation $\hat{U}=U_y^{\dagger} (m) U_x U_y(m+1)U_x^{\dagger}$. }
\end{figure} 
\end{center} 

\item{\it BEC immersion} As mentioned in the text, the immersion method can be used directly to generate the potential Eq.\eqref{gauge}.
We consider atoms with two internals states (two different hyperfine states, or two Zeeman states of the same hyperfine manifold). The rotating BEC creates the Abelian magnetic field with controllable flux $\Phi$ via phonon induced effects as demonstrated in Ref. \cite{imme}. We additionally use the laser assisted tunneling and  lattice tilting methods (and if needed a magnetic field) in order to realize transitions that transfer the used internal states into their superpositions. In this way we create non-trivial hopping operators $U_x,U_y \in U(2)$ and we may control the values of $\alpha$ and $\beta$. In this scheme we note that the lasers no longer need to have a space-dependent phase, since the Abelian part of the gauge potential $\Phi$  is induced by the rotating BEC.

\item{\it Two-photon dressing by laser fields} Finally, the method of dressing by Raman fields \cite{spielman2} could be easily generalized in order to create a non-Abelian vector potential in an optical lattice. The vector potential which is locally created should then depend  on the internal states that are used. 

\end{itemize}

For the concrete realization of the specific gauge potential Eq. \eqref{gauge} that we consider in this work, we may use
$^{40}$K  atoms in  $F=9/2$ or $F=7/2$ hyperfine manifolds, or $^6$Li with $F=1/2$. For $^{40}$K  one should optically pump and restrict the atomic dynamics to the two lowest Zeeman sublevels in each of the 
hyperfine manifolds. One can then employ different lattice tiltings in the $x$ and $y$ directions to perform laser (Raman assisted) tunnelings that change the internal states of the atoms. As mentioned above, this allows to control the parameters $\alpha$ and $\beta$, and fixes the non-Abelian part of the tunneling matrix elelments. In order to generate the Abelian part of the potential,
i.e. the $m$-dependent phase factor in front of $U_y(m)$ in Eq. \eqref{eq3}, there are two possible approaches. One possibility is to follow the lines of Ref. \cite{Osterloh}: the Raman lasers
responsible for the tunneling in the $y$ direction should have a $x$-dependent phase. The other idea is to use the immersion of the system in a rotating BEC \cite{imme}. Both approaches will  allow to control $\Phi$.  In experiments, one routinely reaches the  values of (laser assisted, or direct) tunneling 
rates in the range of 5-10 kHz ($\simeq$ 0.5 $\mu$K), Fermi temperatures  of the same order, and temperatures $T\simeq$ 0.2 $T_{\rm F}\simeq$ 50-100 nK (see for instance \cite{ouradv,Jaksch}).

\section{Gauge symmetry}

In this Appendix we discuss in more details some aspects of the gauge symmetry proper to the models considered in this paper. In particular the distinction between the Abelian and non-Abelian cases is clarified through the introduction of the gauge-invariant Wilson loop.

\paragraph{\it Classical external gauge fields (CEGF)} Let us start by making a clear statement that the gauge fields considered in this paper are classical, external and do not correspond to dynamical variables. The situation is thus analogous to that one encounters in atomic physics when one considers interactions of matter with electromagnetic (EM) fields (i.e. Abelian gauge fields), under the following assumptions: 
\begin{itemize}
\item The EM fields can be regarded as classical, i.e. quantum fluctuations can be neglected. 
\item The EM fields can be regarded as external, and the back influence of the matter on the EM fields is negligible.
\end{itemize}
Such situations typically occur when one deals with strong static electric or magnetic fields, or with strong oscillating fields, like the laser or maser fields. \\
In the present framework, the question of interest is the study of quantum dynamics of matter in such fields. Such situations are also considered in the context of non-Abelian gauge fields, and are widely studied in gauge field theory textbooks \cite{textbook}. Here we  are interested in the quantum dynamics of matter (quarks)
in a given (classical and external) background of the gauge fields. Interesting examples of such fields range from the fields of constant strength, to the fields carrying topological character, such as Polyakov monopoles \cite{textbook}. 

\paragraph{Gauge invariance in Abelian CEGFÕs: the continuum case} We first consider the case of $U(1)$ gauge fields (i.e. for instance EM fields) in the absence of the lattice and we focus on the non-relativistic limit. For our purpose we consider that the gauge potentials give rise to static 
magnetic fields only, and we assume the scalar (electrostatic) potential to be zero (i.e. vanishing electric field).  Generalizations of the following concepts for more general gauge potentials are straightforward.

The second quantized Hamiltonian describing particles subjected to a $U(1)$ CEGF reads: 
\begin{align}
H=& \frac{1}{2m} \int \textrm{d} \bs{x} \, \psi^{\dagger} (\bs{x})  \bigl ( -i \hbar \boldsymbol{\nabla} - \frac{e}{c} \boldsymbol{A}(\bs{x}) \bigr )^2   \psi (\bs{x})  \notag \\
&  + \frac{g}{2} \int \txt{d} \bs{x} \, \psi^{\dagger}(\bs{x})   \psi^{\dagger} (\bs{x})  \psi (\bs{x})   \psi (\bs{x}) ,
\label{ham1}
\end{align}
where $\psi(x)$, $\psi^{\dagger}(x)$ are annihilation and creation operators of the (single component) matter field, $\bs{A}(x)$ is the vector potential, and $g$ is the coupling constant for the contact interactions.  

The Hamiltonian's spectrum is invariant 
with respect to gauge transformations, in which the matter fields undergo a local unitary transformation, 
\be
 \psi (\bs{x})  \rightarrow e^{i \phi (\bs{x})} \psi (\bs{x}) , 
 \label{gtrans}
\ee
and the vector potential transforms according to
\be
\bs{A} (\bs{x}) \rightarrow  \bs{A} (\bs{x}) + \frac{c \hbar}{e} \bs{\nabla} \phi (\bs{x}) .
\label{gtrans2}
\ee

In this context, the magnetic field $\bs{B}= \bs{\nabla} \times \bs{A}$ defines a gauge invariant field. Note that the interaction term in Eq. \eqref{ham1} is invariant under the gauge transformation Eq.\eqref{gtrans}. Generalizations for other kinds of gauge invariant interactions are also straightforward.

\paragraph{Gauge invariance in non-Abelian CEGFÕs: the continuum case}
One can easily extend the above considerations to the situation in which the components of the gauge potential are expressed as non-commutating unitary matrices. The second quantized Hamiltonian describing particles subjected to a $U(N)$ CEGF reads
\begin{align}
H=& \frac{1}{2m} \int \textrm{d} \bs{x} \, \bs{\Psi}^{\dagger} (\bs{x}) \bigl ( -i \hbar \boldsymbol{\nabla} - \frac{e}{c} \boldsymbol{A}(\bs{x}) \bigr )^2  \bs{\Psi} (\bs{x})  \notag \\
&  + \frac{g}{2} \int \txt{d} \bs{x} \, \bs{\Psi}^{\dagger}(\bs{x})   \bs{\Psi}^{\dagger} (\bs{x}) \bs{\Psi} (\bs{x})   \bs{\Psi} (\bs{x}) ,
\end{align}
where $\bs{\Psi} (\bs{x}) $, $\bs{\Psi}^{\dagger}(\bs{x})$ are now annihilation and creation operators of the $N$-component matter fields, $\bs{A}(x)$ is the non-Abelian vector potential, and $g$ is the coupling constant for the contact interactions. The interaction term is now invariant with respect to the $U(N)$ transformations of the matter fields which are described below.

Now, the Hamiltonian's spectrum is invariant 
with respect to $U(N)$ gauge transformations, in which the matter fields undergo a local unitary $U(N)$ transformation  and the vector potential undergoes the gauge change,

\begin{align}
& \bs{\Psi} (\bs{x})   \rightarrow \hat{T} (\bs{x})  \bs{\Psi} (\bs{x}) ,  \\
& \bs{A} (\bs{x}) \rightarrow \hat{T} (\bs{x})  \bs{A} (\bs{x}) \hat{T}^{\dag} (\bs{x})  + \frac{ i c \hbar}{e} \hat{T} (\bs{x}) \bs{\nabla} \hat{T}^{\dag} (\bs{x}) .
\end{align}

The corresponding non-Abelian ``magnetic" field is defined by
\be
 \bs{B} (\bs{x}) = \txt{curl}  \bs{A} (\bs{x})- \frac{i e}{c \hbar} \bs{A} (\bs{x}) \times \bs{A} (\bs{x})  \ee
 and transforms according to 
 \be
  \bs{B} (\bs{x}) \rightarrow \hat{T} (\bs{x})  \bs{B} (\bs{x}) \hat{T}^{\dag} (\bs{x}) .
\ee
Note that the magnetic field no longer describes a gauge invariant quantity in this context.

\paragraph{Gauge invariance in Abelian CEGFÕs: the lattice case} The concept of gauge invariance can also be investigated on lattice systems \cite{munster,rothke}. The Hubbard Hamiltonian that describes particles evolving on a lattice and subjected to a gauge field reads: 
\begin{align}
H=&-t \sum_{\langle i,j \rangle} \biggl [ e^{i \phi_{ij}} b_i^{\dag} b_j +h.c. \biggr ] \notag \\
&+ \frac{V}{2} \sum_i b_i^{\dag}b_i^{\dag}b_i b_i ,
\end{align}
where $b_i$, $b_i^{\dag}$ respectively annihilates and creates a particle on the site $i$, the 
phase factors $\exp(i\phi_{ij})$ are  associated with the tunneling from the nearest neighboring sites $j\to i$, $t$ is the tunneling amplitude and $V$ is the coupling constant for the on-site interactions. Here we consider the case of a two-dimensional square lattice, but the following discussion is valid in general.

In lattice gauge theories, the gauge transformation is naturally defined as an on-site $U(1)$ transformation of the field operators,
\be
b_j \rightarrow b_j'=e^{i \chi_j} b_j.
\label{gtlat}
\ee
The Hamitonian is modified under such a transformation and involves new tunneling phase factors given by
\begin{align}
 \phi_{ij} \rightarrow \phi'_{ij}&=\phi_{ij} + \chi_i -\chi_j , \\
&=\phi_{ij} + \Delta \chi_i . \label{discr}
\end{align}
Note that the latter expression, Eq. \eqref{discr}, is recognized as a discrete form of the gauge transformation Eq. \eqref{gtrans2}.

As in the continuum case, the Hamiltonian's spectrum is left invariant under the gauge transformation Eq. \eqref{gtlat}-\eqref{discr}. When a particle hops around an elementary plaquette, denoted $\square$, it undergoes a unitary transformation $\vert \psi \rangle \rightarrow W(\square) \vert \psi \rangle$, where
 \begin{equation}
W(\square)=e^{i \phi_{1 2}(\square)} e^{i \phi_{2 3}(\square)} e^{i \phi_{3 4}(\square)} e^{i \phi_{4 1}(\square)} ,
\label{uloop}
\end{equation}
and where the plaquette's vertices are situated at neighboring sites $j=1,2,3,4$. This quantity, called the Wilson loop, is therefore associated to the plaquette $\square$ and is defined unambiguously on the lattice by orienting each plaquette in the same way (say clockwise). Obviously the Wilson loops are invariant under the local gauge transformations, Eq. \eqref{discr}. Consequently the set of Wilson loops $\{ W (\square) \}$ determines the physics completely.

Alternatively, the physics is governed by the cumulative phases 
\begin{equation}
\Phi (\square)= \phi_{1 2}(\square)+\phi_{2 3}(\square)+\phi_{3 4}(\square)+\phi_{4 1}(\square) ,
\label{flux}
\end{equation}
that express the Aharonov-Bohm effect: $\Phi (\square)$ is also a gauge invariant quantity and represents the magnetic flux through the plaquette $\square$ in atomic units.  Given a set of Wilson loops $\{ W (\square) \}$ (or fluxes $\{ \Phi (\square) \}$) on the lattice, there exists a choice of phases $\{ \phi_{ij}( \square) \}$, such that Eq. \eqref{uloop} is satisfied globally, i.e. for all loops $\square$. In fact, there exists infinitely many families of local phases, $\{ \phi_{ij}( \square) \}$, that lead to the same physics determined by the given set of Wilson loops, $\{ W (\square) \}$. The latter statement is always valid when open boundary conditions are applied to the lattice. In the case of closed surfaces (such as lattices with periodic boundary conditions, i.e. lattices forming a torus, or symplexes lying on the surface of a sphere) there is an additional physical requirement: the total magnetic flux 
through the closed surface must be zero (in order to verify the Maxwell equation $\bs{\nabla}\cdot \bs{B} = 0$) unless the system presents magnetic monopoles. 
For rigorous proofs of these statements see for instance
\cite{lieb1,lieb2,nachter}.

\paragraph{Gauge invariance in non-Abelian CEGFÕs: the lattice case} 
The Hubbard Hamiltonian that describes particles evolving on a lattice and subjected to a $U(N)$ gauge field reads: 
 \begin{align}
H=&-t \sum_{\langle i,j \rangle} \biggl [ \bs{b}_i^{\dag} \hat{U}_{i j} \bs{b}_j +h.c. \biggr ] \notag \\
&+ \frac{V}{2} \sum_i \bs{b}_i^{\dag} \bs{b}_i^{\dag} \bs{b}_i \bs{b}_i ,
\end{align}
where $\bs{b}_i$, $\bs{b}^{\dag}_i$ are the $N$-component annihilation and creation operators of particles on site $i$, $\hat U_{ij}$ are the 
unitary operators associated with the tunneling from the nearest neighboring sites $j\to i$, and $V$ is the coupling constant for the on-site interactions. 

The on-site gauge transformation is now defined as a $U(N)$ transformation acting on the field operators,
\be
\bs{b}_j \rightarrow \bs{b}_j'=\hat{T}_j \bs{b}_j ,
\label{nagt}
\ee
where $\hat{T}^{\dag} \hat{T} = \hat{T} \hat{T}^{\dag} = \hat{1}$. The transformed Hamitonian  involves new tunneling operators given by
\begin{equation}
\hat U_{ij} \rightarrow \hat T^{\dag}_i \hat U_{ij} \hat T_j .
\label{nagt2}
\end{equation}
The Hamiltonian's spectrum is left invariant under the gauge transformation Eq. \eqref{nagt}-\eqref{nagt2}. Now when a particle hops around an elementary plaquette it undergoes a unitary transformation $\vert \bs{\psi} \rangle \rightarrow \hat{U} (\square) \vert \bs{\psi} \rangle$, where
 \begin{equation}
\hat{U} (\square)= \hat U_{1 2} (\square) \hat U_{2 3} (\square) \hat U_{3 4} (\square) \hat U_{4 1} (\square),
\label{nonabloop}
\end{equation}
and where the plaquette's vertices are situated at neighboring sites $j=1,2,3,4$. This loop operator is associated to the plaquette $\square$ and is defined unambiguously on the lattice by orienting each plaquette in the same way and also by choosing the first link. Physically, the loop operator $\hat{U}$ expresses the non-Abelian Aharonov-Bohm effect \cite{Wilczek,Bohm} which takes place in the system. This important effect accounts for the non-Abelian character of the system and is therefore observable when the loop operator is not reduced to a phase, $\hat{U}= e^{i \theta} \hat{1}$. In the latter trivial case, the non-Abelian Aharonov-Bohm effect cannot take place and the system behaves as an Abelian system.

Under a gauge transformation, the loop operator is modified according to
\be
\hat{U} \rightarrow \hat{T}^{\dagger} \hat{U} \hat{T} ,
\ee
and therefore this operator doesn't  define a gauge invariant quantity. One can then consider the trace of the loop operator,
\be
W(\square)= \txt{tr} \, \hat U_{1 2} (\square) \hat U_{2 3} (\square) \hat U_{3 4} (\square) \hat U_{4 1} (\square) ,
\label{nowil}
\ee
which defines a gauge invariant quantity called the Wilson loop. The set of Wilson loops $\{ W (\square) \}$ therefore determines the physics of the non-Abelian system. \\

Given a set of Wilson loops $\{ W (\square) \}$ on the lattice, there exists a choice of tunneling operators $\{ \hat{U}_{ij}( \square) \}$, such that Eq. \eqref{nowil} is satisfied globally, i.e. for all loops $\square$. Moreover there exists infinitely many families of tunneling operators $\{ \hat{U}_{ij}( \square) \}$ that lead to the same physics determined by the given set of Wilson loops, $\{ W (\square) \}$.

\paragraph{Genuine non-Abelian field configurations} It is now easy to understand which field configuration are genuinely
non-Abelian, and which are not. Let us consider the case of a two-dimensional lattice subjected to a $U(2)$ gauge potential. \\

In the case in which all the loop operators  are trivial and given by complex phase factors, $\hat{U}(\square)=e^{i \theta (\square)} \hat{1}$, we can gauge out all the unitary tunneling operators $\hat{U}_{ij}$, and replace them by simple phase factors. The system then reduces to an Abelian system. For simplicity, let us consider the case in which the loop operators are all equal to the same phase, namely $\hat{U}(\square)=e^{i \theta} \hat{1}$ for all the loops $\square$. We start with the first plaquette and consider the local gauge transformation induced by the following unitary operators:
\begin{align}
& \hat{T}^{\dagger}_1=\hat{1}, \\
& \hat{T}^{\dagger}_2=\hat{U}_{12}, \\
&\hat{T}^{\dagger}_3=\hat{U}_{12}\hat{U}_{23}, \\
&\hat{T}^{\dagger}_4=\hat{U}_{12} \hat{U}_{23} \hat{U}_{34}. 
\end{align}
The resulting new tunneling operators are trivialized according to
\begin{align}
&\hat{U}^{new}_{12}=\hat{T}^{\dag}_1\hat U_{12}\hat T_2=\hat{1}, \\ 
&\hat U^{new}_{23}=\hat{T}^{\dag}_2\hat{U}_{23}\hat{T}_3=\hat{1}, \\
&\hat{U}^{new}_{34}=\hat{T}^{\dag}_3\hat{U}_{34}\hat{T}_4=\hat{1} ,\\
&\hat{U}^{new}_{41}=\hat{T}^{\dag}_4\hat{U}_{41}\hat{T}_1=\hat{U}(\square)=e^{i \theta} \hat{1}.
\end{align}
This trivialization procedure can be extended to the plaquettes that have a common bond with the first one, and so on. Note that if one plaquette $\square_j$ is characterized by a non-trivial loop operator, $\hat{U} (\square_j) \ne e^{i \theta} \hat{1}$, such a gauging out is impossible!

The system studied in this paper features a constant loop operator $\hat{U} (\square)=\hat{U}$, which is explicitely given by Eq. \eqref{Umatrix}. \\
From the above considerations, it follows that when $\hat{U}=\pm e^{i 2 \pi \Phi} \hat{1}$, we can gauge out the unitary tunneling operators  and reduce our problem to two independent Schr\"odinger equations for the two ``flavor" components. If $\hat{U}=e^{i 2 \pi \Phi} \hat{1}$, each ``flavor" component evolves in a square lattice subjected to an Abelian magnetic field, with flux $\Phi$. When  $\hat{U}= -e^{i 2 \pi \Phi} \hat{1}$, the system describes particles evolving on the $\pi$-flux lattice and subjected to an Abelian magnetic field, with flux $\Phi+ \pi$. \\
On the contrary, when the loop operator is non-trivial, $\hat{U} (\square) \ne e^{i \theta} \hat{1}$, no such gauging out is possible and the system cannot be expressed as two independent Abelian models. Moreover, in this case the system exhibits true non-Abelian effects, such as the aforementioned non-Abelian Aharonov-Bohm effect.
The system proposed in this work can therefore be used for non-Abelian interferometry, in which the final effect of an obstacle (a unitary operator acting on the particle's wave function) depends not only on the nature of the obstacle but also on the particle path and location of the obstacle.  \\

In this work, the Abelian and non-Abelian cases are distinguished through a very simple criterion illustrated in Fig. \ref{wilson}: if the Wilson loop's magnitude equals two, $\vert W \vert = \vert \txt{tr} \, \hat{U} \vert =2$, then the loop operator is trivial, namely $\hat{U}= e^{i \theta} \hat{1}$. In this case, we have shown that the system reduces to two independent Abelian systems. In order to apply this criterion, we need to demonstrate the following lemma:

{\it Lemma:} \\
Given a $2 \times 2$ unitary matrix $U \in U(2)$, then
\begin{equation}
\vert \textrm{tr} \, U \vert =2 \iff U= e^{i \theta} \hat{1}
\end{equation}

where tr denotes the trace and $\hat{1}$ is the $2 \times 2$ unit matrix. \\

{\it Proof:} \\
\begin{itemize}

\item If $U= e^{i \theta} \hat{1}$, then of course $\vert \textrm{tr} \, U \vert = \vert e^{i \theta} 2 \vert =2$. \\

\item We now show that if $\vert \textrm{tr} \, U \vert =2$, then $U$ must have the trivial form $U= e^{i \theta} \hat{1}$. Since $U \in  U(2)$, its eigenvalues $\lambda_1=e^{i \theta_1}$ and $\lambda_2=e^{i \theta_2}$ lie on the unit circle and we have that 
\begin{align}
\vert \textrm{tr} \, U \vert &= \vert \textrm{tr} \, T^{\dagger} U  T \vert= \vert \lambda_1 +\lambda_2 \vert = \vert e^{i \theta_1}+e^{i \theta_2} \vert , \\
 &= \sqrt{\bigl ( \cos \theta_1 + \cos \theta_2 \bigr )^2+\bigl ( \sin \theta_1 + \sin \theta_2 \bigr )^2} , \\
& = \sqrt{2 \bigl (  1 + \cos (\theta_1 - \theta_2)   \bigr )} , 
\end{align}
where the unitary matrix $T \in U(2)$ diagonalizes the matrix $U$. 
Therefore the hypothesis $\vert \textrm{tr} \, U \vert =2$ is satisfied if 
\begin{equation}
\theta_1 - \theta_2 = 2 \pi n ,
\end{equation}
 where $n$ is an integer. Under this condition, $T^{\dagger} U  T = \textrm{diag} (\lambda_1) =e^{i \theta} \hat{1}$, where $e^{i \theta}=\lambda_1 = \lambda_2$. Eventually this shows that $U= e^{i \theta} \hat{1}$, since $T T^{\dagger}=T^{\dagger} T=\hat{1}$.  $\square$
\end{itemize}



\begin{thebibliography}{99}

\bibitem{Bloch}I. Bloch, J. Dalibard, and W. Zwerger, Rev. Mod. Phys. {\bf 80}, 885 (2008)
\bibitem{ouradv}M. Lewenstein, A. Sanpera, V. Ahufinger, B. Damski, A. Sen(de) and U. Sen, Adv. Phys. {\bf 56}, 243 (2007).

\bibitem{Ho2001} T.-L. Ho, Phys. Rev. Lett. {\bf87}, 060403 (2001).



\bibitem{Jaksch}D. Jaksch and P. Zoller, New J. Phys. {\bf 5}, 56 (2003).

\bibitem{Mueller}E. J. Mueller, Phys. Rev. A {\bf70}, 041603(R) (2004).

\bibitem{Demler}A.S. S{\o}rensen, E. Demler, and M.D. Lukin, Phys. Rev. Lett. {\bf94}, 086803 (2005).

\bibitem{Ohberg}G. Juzeli{\= u}nas, P. \"Ohberg, J. Ruseckas and A. Klein, Phys. Rev. A {\bf 71 },  053614 (2005).

\bibitem{Ohberg2}G. Juzeli{\= u}nas and P.  \"Ohberg, Phys. Rev. Lett. {\bf93}, 033602 (2004).

\bibitem{spielman} For the first experiments along these lines see: Y.-J. Lin, W.D. Phillips, J.V. Porto, and I.B. Spielman, Bull. Am. Phys. Soc. {\bf 53}, No. 2, A14.00001 (2008).

\bibitem{spielman2} Y.-J. Lin, R.L. Compton, A.R. Perry, W.D. Phillips, J.V. Porto, and I.B. Spielman, arXiv:0809.2976v1 [cond-mat.other].

\bibitem{Holland1}R. Bhat, L. D. Carr and M. J. Holland, Phys. Rev. Lett. {\bf96}, 060405 (2006).

\bibitem{Polini2005} M. Polini, R. Fazio, A. H. MacDonald and M. P. Tosi,  Phys. Rev. Lett. {\bf 95}, 010401 (2005)

\bibitem{Holland2}R. Bhat, M. Kr\"amer, J. Cooper and M. J. Holland,  Phys. Rev. A {\bf76}, 043601 (2007).

\bibitem{Tung} S. Tung, V. Schweikhard, and E. A. Cornell, Phys. Rev. Lett. {\bf97}, 240402 (2006).

\bibitem{imme} A. Klein and D. Jaksch, Europhys. Lett. {\bf 85}, 13001 (2009).


\bibitem{Hofstadter}D. Hofstadter, Phys. Rev. B {\bf 14}, 2239 (1976).

\bibitem{Goldman2}N. Goldman, Europhys. Lett. {\bf 80}, 20001 (2007). 

\bibitem{Palmer} R.N. Palmer and D. Jaksch, Phys. Rev. Lett. {\bf96}, 180407 (2006).

\bibitem{Goldman1}N. Goldman and P. Gaspard, Europhys. Lett. {\bf 78}, 60001 (2007). 

\bibitem{Hafezi}M. Hafezi, A. S. S\o rensen, M. D. Lukin and E. Demler, Europhys. Lett.  {\bf 81}, 10005 (2008).

\bibitem{Osterloh}K. Osterloh, M. Baig, L. Santos, P. Zoller and M. Lewenstein,  Phys. Rev. Lett. {\bf 95}, 010403 (2005).

\bibitem{juze} J. Ruseckas, G. Juzeli{\= u}nas, P. \"Ohberg, and M. Fleischhauer, Phys. Rev. Lett. {\bf 95}, 010404 (2005).


\bibitem{Clark}I.I. Satija, D. C. Dakin, J. Y. Vaishnav and C. W. Clark,  Phys. Rev. A {\bf 77}, 043410 (2008).

\bibitem{santos} A. Jacob, P. \"Ohberg, G. Juzeli{\= u}nas and L. Santos, New J. Phys. {\bf 10} 045022 (2008).

\bibitem{santos2} A. Jacob, P. \"Ohberg, G. Juzeli{\= u}nas and L. Santos, Appl. Phys. B {\bf 89}, 439-445 (2007).

\bibitem{santos3} G. Juzeli{\= u}nas, J. Ruseckas, A. Jacob, L. Santos and P. \"Ohberg, Phys. Rev. Lett. (in press), arXiv:0801.2056.

\bibitem{Umu} R. O. Umucalilar, Hui Zhai, and M. \"O. Oktel, Phys. Rev. Lett. {\bf 100}, 070402 (2008).

\bibitem{rem}{Close to the Abelian regime, the Hall conductivity can be explicitly computed as some large gaps persist inside the ``moth" spectrum \cite{Goldman1}.}

\bibitem{Travel} M. Lewenstein, A. Kubasiak, J. Larson, C. Menotti, G. Morigi, K. Osterloh and A. Sanpera,  Proc. of ICAP-2006 Innsbruck, Eds. C. Roos and H. Haeffnerand R. Blatt, (AIP, Melville, NY, 2006), 201-212.
 
\bibitem{remark}{Chern numbers \cite{Nakahara} have been used in the IQHE context \cite{Thoules}, as well as to characterize strongly correlated states \cite{Wen1,Hafezi1}. The Chern analysis was performed in Refs. \cite{Goldman1,Goldman3,Goldman4} in order to characterize the IQHE in atomic systems and quantum graphs.}

\bibitem{graphene} Y. Hatsugai, T. Fukui, and H. Aoki, Phys. Rev. B {\bf 74}, 205414 (2006).

\bibitem{Thoules} D. J. Thouless, M. Kohmoto, M. P. Nightingale, and M. den Nijs, Phys. Rev. Lett. {\bf 49}, 405 (1982).

\bibitem{Nakahara}M. Nakahara, {\it Geometry, Topology and Physics}, 2nd edition (Institute of Physics, Bristol, 2003).

\bibitem{Fukui}T. Fukui, Y. Hatsugai, and H. Suzuki, J. Phys. Soc. Jap. \textbf{74}, 1674 (2005).

\bibitem{Osadchy}D. Osadchy and J. Avron, J. Math. Phys. \textbf{42}, 5665 (2001).

\bibitem{dioremark} The quantized conductivity depends on the real-valued parameters $\alpha$ and $\beta$. Thus a Diophantine equation cannot be established here, contrary to the Abelian case \cite{Thoules}.

\bibitem{Wen1}X. G. Wen and Q. Niu, Phys. Rev. B {\bf 41}, 9377 (1990).

\bibitem{Hafezi1} M. Hafezi, A. S. S\o rensen, E. Demler and M. D. Lukin, Phys. Rev. A {\bf 76}, 023613 (2007).

\bibitem{Goldman3}N. Goldman and P. Gaspard,  Phys. Rev. B {\bf 77} 024302 (2008).

\bibitem{Goldman4}N. Goldman,  J. Phys. B: At. Mol. Opt. Phys. {\bf 42}  055302 (2009)

\bibitem{munster} I.Montvay and G. M\"unster, {\it Quantum Fields on a Lattice (Cambridge Monographs on Mathematical Physics)} (Cambridge University Press, Cambridge, 1997).

\bibitem{rothke} H.J. Rothe, {\it Lattice Gauge Theories: An Introduction (World Scientific Lecture Notes in Physics)}, (World Scientific, Singapore, 2005).

\bibitem{textbook} V. Rubakov and S.S. Wilson,  {\it Classical Theory of Gauge Fields}, (Princeton University Press, Princeton, 2002)

\bibitem{lieb1} E.H. Lieb, Phys. Rev. Lett. {\bf 73}, 2158 (1994).

\bibitem{lieb2} E.H. Lieb, Helv. Phys. Acta {\bf 65}, 247 (1992).

\bibitem{nachter} N. Macris and B. Nachterg\"ale, J. Phys. A {\bf 85}, 745 (1996)

\bibitem{Wilczek} F. Wilczek, A. Shapere,  {\it Geometric Phases in Physics}, (World Scientific, 1989)

\bibitem{Bohm} A. Bohm, A. Mostafazadeh, H. Koizumi, Q. Niu, J. Zwanziger,  {\it The Geometric Phase in Quantum Systems}, (Springer, 2003)


\end{thebibliography}
\end{document}